\documentclass[universe,article,accept,pdftex,moreauthors]{Definitions/mdpi} 
\firstpage{1} 
\pubvolume{1}
\issuenum{1}
\articlenumber{0}
\pubyear{2022}
\copyrightyear{2022}
\datereceived{} 
\dateaccepted{} 
\datepublished{} 
\hreflink{https://doi.org/} 

\Title{Testing the Wave-Particle Duality of Gravitational Wave Using the Spin-Orbital-Hall Effect of Structured Light}

\TitleCitation{Test the Wave-Particle Duality of Gravitational Wave Using the Spin-Orbital-Hall Effect of Structured Light}

\Author{Qianfan Wu $^{1}$, Weishan  Zhu$^{1}$\orcidB{} and Longlong Feng$^{1, *}$\orcidA{}}

\AuthorNames{Qianfan, Wu, Weishan Zhu and Longlong Feng}

\AuthorCitation{Wu, Q.F.; Zhu, W.S.; Feng, L.L.}

\address{%
$^{1}$ \quad School of Physics and Astronomy, Sun Yat-Sen University, Zhuhai 519082, China;  Wu: wuqf6@mail2.sysu.edu.cn; Zhu: zhuwshan5@mail.sysu.edu.cn; Feng: flonglong@mail.sysu.edu.cn}

\corres{Correspondence: flonglong@mail.sysu.edu.cn}

\abstract{Probing the polarization of gravitational waves (GWs) would provide an evidence of graviton, indicating the quantization of gravity. Motivated by the next generation of gravitational wave detectors, we make an attempt to study the possible helicity coupling of structured lights to GWs.  With the analogue between of gravitational fields and the generic electromagnetic media, we present a 4-vector optical Dirac equation based on the Maxwell theory under the paraxial approximation. It is found that twisted lights propagating in a gravitational field can be viewed as a non-Hermitian system with the $\mathcal{PT}$ symmetry. We further demonstrate that the coupling effect between angular momentums of the GWs and twisted lights may make photons undergo both dipole and quadrupole transitions between different orbital-angular-momentum(OAM) eigenstates and leads to some measurable optical features, including the central intensity brightening and macroscopic rotation of the intensity pattern for twisted lights. The former is spin-independent while the later is spin dependent phenomena, both of which can be viewed alternatively as the spin-orbital-Hall effect of structured lights in the GWs and can serve as an indicator of the particle nature of GWs.}


\begin{document}


\section{Introduction}

The detection of gravitational wave (GW) by the LIGO interferometer fulfilled experimental verification of the last prediction from the linearized Einstein theory of general relativity \cite{PhysRevLett.116.061102, PhysRevLett.116.221101}.  In the low energy regime, the canonical quantization of linear perturbations around the Minkowski background yields a massless spin-2 particle, called graviton\cite{PhysRev.138.B988}. Though we believe that the long-range gravity is mediated by the graviton, its existence has never been justified experimentally due to its extraordinary weakness. Actually, measuring graviton is to make sense of its particle nature of GWs according to wave-particle duality at the quantum level, namely, recognizing some particle features by direct or indirect detections\cite{10.1103/physrevd.104.046021}.  One theoretical possibility was suggested to detect a single graviton by the emission and absorption of gravitons\cite{10.1088/0264-9381/23/20/006, 10.1007/s10701-006-9081-9}, and has been revisited by considering the transition between discrete quantum states with different angular momentums\cite{10.1103/physrevd.104.065016}. 
  
Moreover, the direct detection of gravitational waves by the advanced LIGO interferometer and the advanced Virgo interferometer marks the beginning of the era of gravitational wave astronomy \cite{PhysRevLett.116.241103, PhysRevX.6.041015, PhysRevLett.118.221101}. These discoveries stimulate the great efforts devoted to develop more advanced optical detection technology. With upgrading to next generation of interferometer,  it is expected to move to design sensitivity and observe ever-increasing numbers of GW sources.  The next generation of GW laser-interferometric GW detector will be upgraded to have an unprecedented sensitivity by minimizing various technical noises. One limiting noise arises from mirror thermal noise, producing by Brownian motion of particles in coatings and substrates.  Besides the physically cooling of mirror, an alternative solution is to change the mode shape of the laser beam inside the interferometer, e.g. resonating the higher-order Laguerre Gauss (LG$_p^l$) modes in the detector arm cavities to smooth out the thermal noise fluctuations over a bigger portion of the mirror surface\cite{PhysRevD.79.122002}.  A well-developed technology has made it possible to produce higher-order LG-modes with high power output and high mode purity as required in the GW detection\cite{PhysRevLett.105.231102, PhysRevLett.110.251101}. Therefore, it would be crucial to understand the wave mechanics of the higher-order LG modes interacting with the passing gravitational wave. 

Though the wave theory offers a complete classical description of optical phenomena in nature, in most application situations, the typical wavelength of light is much less than the inhomogeneous length scale of medium, the geometric optics approximation is valid and the correspondence principle allows us to make a simple analysis in context of the Hamiltonian dynamics of particle, which can be induced from the dispersion relation associated with the wave equation. On the other hand, the Maxwell theory encodes intrinsic degrees of freedom - the polarizations associated with rotating electric and magnetic fields, corresponding to the two spin states of photons. In addition, lights can possess two types of orbital angular momentums (OAM) - the intrinsic OAM from 
twisted wavefront of structured lights, and the extrinsic orbital angular momentum from helical optical paths in inhomogeneous media. It has been realized that the interplay between various optical OAM - spin-orbit interaction of light (SOI),  will lead to a variety of optical phenomena on subwavelength scales\cite{ 10.1038/nphoton.2015.201,10.1016/j.physrep.2015.06.003, 10.1038/nphoton.2015.203, 10.1103/physrevlett.126.243601}. The important manifestation of the SOI phenomena is the spin-Hall effects in inhomogeneous media arising from the spin-extrinsic OAM interaction, indicating how the intrinsic polarization degrees of freedom affect the light trajectories beyond the geometric optics approximation\cite{10.1103/physrevlett.93.083901, 10.1103/physreva.75.053821, 10.1103/physreva.46.5199, 10.1038/nphoton.2008.229, 10.1088/1464-4258/11/9/094009, 10.1016/j.physleta.2004.10.035, 10.1103/physrevd.74.021701, 10.1103/physreva.92.043805}. 

For light propagation in a gravitational field, the curved space background can be transformed to a linear optical medium, whose optical properties characterized by the effective dielectric tensor are fully specified by the spacetime geometries. A typical example is the deflection of light in gravitational fields, this phenomenon can be illustrated by the correspondence between the refractive index and the scalar gravitational potential. Again, a rotating gravitational field exhibits the optical activity attributed to the vortical dragging vector, corresponding to gyrotropic materials\cite{10.1103/physrevd.46.5407, doi:10.1142/S0218271801001402, MASHHOON1993347, andp.20035150501}. By this analogue, the gravitational spin-Hall effect can be easily understood for lights propagating in curved spacetime \cite{10.1103/physrevd.102.024075, 10.1038/nphys1907, 10.1103/physrevd.104.084007, 10.1103/physreva.104.013718}. 

The motivation of this work is two-fold. Firstly, the next generation of GW detectors may provide more information about the polarization modes of GWs, This paper is to explore the spin-orbital-Hall effect of structured light due to GWs, which may potentially reveal more polarization structures of GWs and allows for testing the theory of gravity beyond GR and justifying the particle nature of GWs. Secondly, it would be interesting to seek for an alternative mechanism for detecting GWs in the polarization space, which requires a full vector-wave analysis of structured light coupling with GWs on subwavelength scales.    

However, It is noted that the current approach for the spin-Hall effect of light is based on the geometrodynamics of spinning particles\cite{10.1103/physreva.46.5199, 10.1038/nphoton.2008.229, 10.1088/1464-4258/11/9/094009, 10.1016/j.physleta.2004.10.035, 10.1103/physrevd.74.021701}, which is not applicable to dealing with structured lights with intrinsic OAMs propagating in an inhomogeneous medium. In this case, a full vectorial-wave mechanics is required to describe the behavior of polarized structured lights. In this work, based on the four-vector optical Dirac equation developed by Feng \& Wu \cite{2022arXiv220314664F}, we investigate the spin-Hall effect of light in GWs, seeking for an alternative method to detect GWs. The paper is organized as follows.  The optical Dirac equation of paraxial light in linearized gravitational fields is presented in section 2. We further discuss the dipole and quadrupole interactions of structured light with GWs in section 3. In section 4, we present perturbation analyses of the GWs' signals in two numerical experiments and discuss theirs physical implications. Finally we summarize the paper and give concluding remarks in section 5.  

\section{Optical Dirac Equation of Paraxial Light in Gravitational Fields}

Let us adopt the following 3+1 decomposition, e.g. \cite{10.1103/physrevd.16.933}, defining the spatial metric by 
\begin{equation}
\gamma_{ij}=g_{ij}, \quad \gamma^{ij}= g^{ij} - g^{0i}g^{0j}/g^{00}
\end{equation}
and the 3d dragging vector ${\bf g}$ as 
\begin{equation}
g_i=-g_{0i}, \quad g^i=g^{0i}/g^{00}, \quad
 \end{equation}
The electromagnetic vectors are identified by 
\begin{eqnarray} 
&E_i= F_{i0}, \quad &B^i=\frac{1}{2\sqrt{\gamma}}\epsilon^{ijk}F_{jk} \\
&D^i=F^{0i}/\mathcal{N}, \quad &H_i=\frac{1}{2}\sqrt{\gamma}\epsilon_{ijk}F^{jk}/\mathcal{N}
\end{eqnarray}
where $\mathcal{N}=\sqrt{-g^{00}}$, $\gamma$ is the determinant of $\gamma_{ij}$. These definitions lead us to the Maxwell equation in a noncovariant form
\begin{eqnarray}
\nabla \times \mathbf{E}=-\frac{(\sqrt{\gamma} \mathbf{B})_t}{\sqrt{\gamma}}, & \nabla \cdot \mathbf{B}=0 \\
\nabla \times \mathbf{H}=\frac{(\sqrt{\gamma} \mathbf{D})_t}{\sqrt{\gamma}}, & \nabla \cdot \mathbf{D}=0
\end{eqnarray}
in which, ${\bf D}$, ${\bf B}$ are related to ${\bf E}$, ${\bf H}$ by the constitutive equations,
\begin{eqnarray} 
{\bf D} &= \mathcal{N}({\bf E}+{\bf g}\times{\bf B}) \\
{\bf B} &= \mathcal{N}({\bf H}+{\bf D}\times{\bf g}) 
\end{eqnarray}

In a linearized gravitational wave theory, $g_{\mu\nu} = \eta_{\mu\nu} + h_{\mu\nu}$, $|h_{\mu\nu}| \ll 1$, the space-time background can be approximated adiabatically by an irrotational gravitational field, thus, $\mathcal{N}=1$, ${\bf g} = 0$, the electric field  ${\bf E}$ and magnetic field ${\bf B}$ are related to the auxiliary fields ${\bf D}$ and ${\bf H}$ simply by the local constitutive equation $D^i = \gamma^{ij}E_j$, $B^i= \gamma^{ij}H_j$. Formally, for a light beam propagating in gravitational wave, the gravitational field is equivalent to an optical medium with the permittivity and permeability tensors obeying a simple equality relation with the spatial metric $\gamma = \{g_{ij}\}$, 
$\epsilon^{-1} = \mu^{-1}=\gamma$, which actually corresponds to the spin-degenerate condition in photonic topological insulators\cite{10.1038/nmat3520}. 
Since the wavelength of light is much less than that of GW, the GW is further assumed to be a static gravitational field. 
Define a 6-vector, $ {\mathcal F} =({\bf D}, i{\bf B})^{T}$, in this way, the Maxwell equation becomes a Schr\"{o}dinger-like compact form
\begin{equation}\label{Maxwell}
i\frac{\partial {\mathcal F}}{\partial t} = \sigma_{x}\otimes [({\bf k}\cdot{\bf s})\cdot {\gamma}] {\mathcal F}
\end{equation}

As having been discussed by Feng \& Wu\cite{2022arXiv220314664F}, the Maxwell equation in generic media can be converted to a non-Hermitian $\gamma_5$ extended Dirac theory for massive fermions. Central to this approach is to perform vectorial analysis in the helicity basis spanned by the transverse basis ${\bf e}_{\pm}$ with respect to a given direction, which can be aligned with propagation direction of incident light, e.g. without loss of generalities, taking the z-axis. In this case, the helicity basis satisfies $({\bf e}_z\cdot{\bf s}){\bf e}_{\pm} = \pm {\bf e}_{\pm}$. The transversality condition places a constraint on the polarization states of photons, implying the field equation Eq.(\ref{Maxwell}) is reducible.  It is noted that, under the paraxial approximation, the longitudinal components can be derived explicitly from the transverse condition, and the electromagnetic 6-vector ${\mathcal F}$ can be reduced to four independent components. Introducing the vector wavefunctions
\begin{equation}\label{WF}
{\bf \Psi}_{\pm} = {\mathcal D}_{\perp}\pm i\sigma_3 {\mathcal B}_{\perp}
\end{equation}
with the transverse components$\mathcal{D}_{\perp} = (D_{+}, D_{-})^{T}$ and $\mathcal{B}_{\perp} = (B_{+}, B_{-})^{T}$ in the helicity space, and combining ${\bf \Psi}_+$ and ${\bf \Psi}_-$ to form a 4-vector ${\bf \Psi} = ({\bf \Psi}_+, {\bf \Psi}_-)^T$
we can arrive at the following optical Dirac equation,
\begin{equation}\label{Dirac}
i\frac{\partial}{\partial t} {\bf \Psi} = (\beta \hat{m}  + \beta {\alpha} \cdot \hat{\bf p}){\bf \Psi}
\end{equation}
in which the effective mass term is 
\begin{equation}\label{EM}
\hat{m} = (1+Q_0)\hat{k}_z + \frac{1+q_0}{2k_z}\hat{\bf k}_{\perp}^2 - {\bf q}\cdot\hat{\bf k}_{\perp}
\end{equation}
and the momentum components are given in the helicity basis through $\hat{\bf p} \cdot{\sigma} = {\mathrm p_+} \sigma_{+} + {\mathrm p_-}\sigma_{-} + p_z \sigma_z$ and $\sigma_{\pm} = \frac{1}{2}(\sigma_x \pm i \sigma_y)$, 
\begin{equation}\label{EP}
{\mathrm p}_{\pm}= Q_{\pm 2}\hat{k}_z  - 2Q_{\pm 1}\hat{k}_{\pm}  + \frac{1+q_0}{k_z}\hat{k}_{\pm}^2, \quad p_z=0,
\end{equation} 
where $Q_0 \equiv \frac{1}{2}(h_{11}+h_{22})$ and $q_0=h_{33}$ are the monopoles, ${\bf q} = \{Q_{+1}, Q_{-1}\}^{T}$ is the dipole momentum with the components $Q_{\pm 1}=(h_{13}\mp ih_{23})/\sqrt{2}$,  the quadrupoles $Q_{\pm 2} = \frac{1}{2}(h_{11} - h_{22}) \mp ih_{12}$. The effective momentums Eq.(\ref{EP}) only have surface terms in the transverse plane. It can be seen that Eq.(\ref{Dirac}) describes a non-Hermitian system with the PT symmetry\cite{10.1088/0034-4885/70/6/r03, 10.1080/00018732.2021.1876991, 10.1038/nphys4323}, and is similar to the Jackiw-Rebbi model in topological insulators\cite{10.1007/978-3-642-32858-9, 10.1103/physreva.100.053819}. The only difference between the optical Dirac equation Eq.(\ref{Dirac}) and the Dirac equation for spin $1/2$ particle is an extra $\beta$ operation appearing on the momentum term in the former. 

Under the geometric optics approximation, keeping the leading term $\sim O(k_z)$ in Eq.(\ref{Dirac}) yields the energy-eigenvalue equation
\begin{equation}\label{EN}
\lambda^{2} + |Q_{+2}|^2 = (1+Q_0)^2
\end{equation}
where $\omega = \lambda k_z$. Eq.(\ref{EN}) has two solutions with the positive and negative energies. Since the anti-particle of a photon is itself, a photon with negative energy and helicity can be regarded as a mirror photon with positive energy and opposite helicity. According to Eq.(\ref{EN}), its solution can be parameterized by $\lambda = (1+Q_0) \cos\theta_Q$, $Q_{\pm 2} = |Q_{\pm 2}| e^{\pm\phi_Q} = (1+Q_0)\sin\theta_Q e^{\pm\phi_Q}$. 
Accordingly, the positive energy eigenstates with the opposite helicities are written as
\begin{equation}\label{ES}
|+\rangle=\left(\begin{array}{c}
\cos\displaystyle{\frac{\theta_Q}{2}} \\
0 \\
0 \\
-\sin\displaystyle{\frac{\theta_Q}{2}} e^{-i\phi_Q}
\end{array}\right) ; 
|-\rangle=\left(\begin{array}{c}
0 \\
\cos\displaystyle{\frac{\theta_Q}{2}} \\
-\sin\displaystyle{\frac{\theta_Q}{2}} e^{i\phi_Q}\\
0
\end{array}\right)
\end{equation}
We consider the positive-energy solutions to Eq.(\ref{Dirac}) only, which can be written by a linear superposition of the two helicity eigenstates Eq.(\ref{ES}),  
\begin{equation}
{\bf \Psi} = (\varphi_{+} |+\rangle + \varphi_{-}|-\rangle){\mathrm e}^{-i\omega t + ik_z z}
\end{equation}
Define 2-spinor ${\Phi} = (\varphi_{+}, \varphi_{-})^{T}$, we obtain the optical Schrodinger equation
\begin{equation}\label{Paxa}
i(1+Q_0)\frac{\partial \Phi}{\partial z} = -\frac{1+q_0}{2k_z}\nabla^2_{\perp} \Phi  + V({\bf r}_{\perp})\Phi
\end{equation}
with the complex optical potential
\begin{equation}\label{PH}
V({\bf r}_{\perp}) =  - {\bf q} \cdot {\bf k}_{\perp} + \sigma_3  \frac{1+q_0}{2\lambda k_z}\bigl[ Q_{-2}\hat{k}_{+}^2 - Q_{+2}\hat{k}_{-}^2\bigr]
\end{equation}
where only the terms of linear order of $O(h_{ij})$ are kept as in the linearized gravity theory. Clearly, above equation takes a similar form as the simplest optical model describing the paraxial propagation of light in an inhomogeneous dielectric medium with a complex refractive index\cite{PhysRevLett.100.103904, 10.1002/lpor.200810055}, whereas here is emulating a quantum two-component system. The potential Eq.(\ref{PH}) consists of two parts, the real part gives a dipole interaction, while the imaginary part is a non-Hermitian quadrupole interaction corresponding to the optical gain or loss of the medium. On the other hand, the Hamiltonian Eq.(\ref{Paxa}) has been actually decoupled into two independent scalar equations, implying that the spin-up and spin-down state evolve independently and thereby no helicity transition occurs.  The helicity dependent quadrupole coupling is anti-Hermitian conjugate each other for the spin-up and -down states, which ensures the $\mathcal{PT}$ symmetry of the optical fields and leads to the optical chirality.

In a flat space, the optical potential in the right side of Eq.(\ref{Paxa}) due to the spacetime perturbations vanished, and thus Eq.(\ref{Paxa}) reduces to the familiar paraxial equation, which could have family solutions of structured light with OAMs. In the cylindrical coordinates, the paraxial equation has a set of solutions of the Laguerre-Gaussian modes 
\begin{eqnarray}\label{LG}
LG_{n}^{l}(\rho, z) & =& \frac{a_n^l}{w(z)}\Bigl(\frac{\sqrt{2} \rho}{w(z)}\Bigr)^{\vert l \vert}L_{n}^{\vert l \vert}\Bigl(\frac{2\rho^2}{w(z)^2}\Bigr) \exp{\Bigl[-\frac{\rho^2}{w(z)^2}\Bigr]} \nonumber \\
& \cdot &\exp{\Bigl(i\frac{k\rho^2}{2R}\Bigr)}\exp{(il\phi)}\exp{(-i\varphi(z))} 
\end{eqnarray}
where $n$ and $l$ are the radial and the azimuthal indices, the order of the model is given by $N= 2n+|l|$, the constant $a_n^l = (2 n!/\pi(n+|l|)!)^{1/2}$,  $w(z)$ is the width of mode, $R$ the radius of the wavefront curvature, and the Gouy phase factor $\varphi(z) = (N+1)\tan^{-1}(z/z_R)$, $z_R = \frac{1}{2}kw_0^2$, here $w_0=w(z=0)$ is the beam waist.

\section{Dipole and Quadrupole Interaction of Photons and Gravitational Waves}

In a metric gravity theory, GWs are allowed, at most, to have six polarization modes, including two tensor-types(spin-2), two vector types(spin-1) and two scalar-types (spin-0)\cite{10.1103/physrevlett.30.884, will2018theory}. The dipole momentum ${\bf q}$ in Eq.(\ref{PH}), $H_{D} = - {\bf q}\cdot{\bf k}_{\perp}$ is associated with the vector-type excitation. If the incident vector ${\bf e_g}$ of GWs is at a given direction of $\{\theta_g, \phi_g\}$ in the spherical coordinates with respect to the propagating direction ${\bf e}_k $ of the light beam, the Einstein theory only allows for the existence of the tensor polarizations of plus $(+)$ and cross$(\times)$,     
\begin{equation}\label{qvec}
{\bf q} = h^{+}({\bf e}_g\cdot{\bf e}_k)[({\bf e}_g\cdot{\bf e}_k){\bf e}_k-{\bf e}_g]+h^{\times}({\bf e}_g\times{\bf e}_k)
\end{equation}
In case of the incident GW propagating parallel with the light beam, the dipole is identically zero, ${\bf q} =0$, as given in the TT gauge. This dipole interaction  consists of the two contributions, one is $\propto  h^{+}({\bf e}_g\cdot{\bf k}_{\perp})$, another is 
$\propto h^{\times}[{\bf e}_g\cdot({\bf e}_k\times{\bf k}_{\perp})]$. Formally, the former is from the GW polarization coupled with the OAM density flow of photons ${\bf S}_o \propto \mathrm{Im}[{\bf E}^*\cdot(\nabla {\bf E})]$  , and the latter from the spin density flow ${\bf S}_c \propto \mathrm{Im}[\nabla\times({\bf E}^{*}\times{\bf E})]$\cite{10.1088/1464-4258/11/9/094001}. 

In the "helicity basis", the dipole interaction can be alternatively written by
\begin{equation}\label{Qint}
H_{D} = i\bigl[Q_{-1}\nabla_{+}+Q_{+1}\nabla_{-}\bigr]
\end{equation}
where
$$ iQ_{\pm} = -\frac{1}{\sqrt{2}}\sin\theta_g(h^{\times}\pm i\cos\theta_gh^{+})e^{-i\phi_g}$$
The complex differential operators $\nabla_{\pm}$ in the cylindrical coordinates become
\begin{equation}\label{diff}
\nabla_{\pm} = \frac{1}{\sqrt{2}}({\bf \nabla}_x \mp i{\bf \nabla}_y)=\frac{1}{\sqrt{2}}e^{\mp i\phi}(\partial_{\rho} \pm \frac{1}{\rho}L_z)
\end{equation}
where $L_z= -i\partial_\phi$ is the orbital angular momentum operator in z-direction.  It is noted that, for the eigenstates $L_z \vert l\rangle = l \vert l\rangle$, 
${\bf \nabla}_{\pm}$ act as the ladder operators to lowers/raises OAMs by one unit.  Thus, the dipole interaction will make the initial $l$ mode to produce two extra modes with the OAM number of $l-1$ and $l+1$. For the LG modes, $\vert n,  l \rangle = LG_{n}^{l}(\rho, z)$, it is not difficult to work out the following ladder relations for the lowering and raising operators, 
\begin{eqnarray}
\label{ladder1}&\nabla_{\pm}\vert n,  \pm l \rangle =  k_w\bigl[\sqrt{n+l}\vert n,\pm(l-1)\rangle+\sqrt{n+1}\vert n+1,\pm(l-1)\rangle\bigr]  \\
\label{ladder2}&\nabla_{\mp}\vert n, \pm l \rangle = k_w\bigl[\sqrt{n+l+1}\vert n,\pm(l+1)\rangle - \sqrt{n}\vert n-1,\pm(l+1)\rangle\bigr]
\end{eqnarray}
where $k_w = 1/w_0$, $l>0$ and $\pm$ signs are taken meantime on both sides of the equations. 

The imaginary part in the complex potential Eq.(\ref{PH}) is helicity dependent quadrupole interaction. In the Einstein gravity, the quadrupole moment can be  expressed explicitly as 
 \begin{equation}
Q_{\pm 2}=\bigl[ \frac{1}{2} h^{+}(1+\cos^2\theta_g) \mp ih^{\times}\cos\theta_g\bigr]e^{\mp i2\phi_g}
\end{equation}
which are two tensor types associated with the spin-2 excitations, i.e., gravitons. Physically, this quadrupole interaction could be understood as the coupling effect between the helicity of GWs and OAM of twisted lights, which will cause photons to undergo quadrupole transitions by decreasing and increasing the OAMs of two quantum units simultaneously. For the quadrupole coupling $Q_{-2}\nabla_{+}^2$, the photon will lose two units of OAMs by emitting a graviton, and conversely,  the coupling $Q_{+2}\nabla_{-}^2$ will cause the photon gain two units of OAMs through absorbing a graviton. Due to the helicity dependent of the quadrupole interaction, the resulting extra helicity modes would have opposite signs for right and left circularly polarized lights, i.e., an overall phase difference of $\pi$. Actually, this chirality is due to the 'time-reversal symmetry breaking'\cite{10.1088/2040-8986/aaaa56}. Recall that, in the classical electromagnetism, this breaking is always attributed to the axial property of the magnetic field vector, i.e., even in $\mathcal{P}$ 
and odd in $\mathcal{T}$, and is manifested by a classical example of the Faraday rotation, in which, the polarization plane of a light beam will acquire a rotation angle in a magnetic field aligned with the propagating direction of lights. As we will demonstrated later, this quadrupole coupling effect will also exhibit a new optical activity for polarized twisted lights.  

In addition, it should be emphasized here that there exists an addition rotation freedom from the GW polarization around ${\bf e}_g$, with respect to which the polarizations $\{h_{+},h_{\times}\}$ are defined.  Performing an rotation in the transverse place ${\bf e}_{\perp}' = R(\psi){\bf e}_{\perp}$, (R is a rotation transformation with an angle $\psi$),  we have $\{h'_{+},h'_{\times}\}^{T} = R(2\psi)\{h_{+},h_{\times}\}^{T}$. 

\section{Perturbation Analysis and Numerical Experiments}

In the following, we will present two typical ideal experiments to demonstrate the optical features induced by the dipole and quadrupole interaction between structured lights and gravitational waves.  We can apply the perturbation theory to solve Eqn.(\ref{Paxa}). Since the Laguerre-Gaussian modes form a complete and orthonormal set with respect to the mode indices $n$ and $l$ in the polar plane $\{\rho, \phi\}$, we can make a decomposition such that 
\begin{equation}\label{LGseries}
\varphi_{\pm}(z)= \sum_{m,k} \xi_{m.k}^{\pm}(z) \vert m,k\rangle
\end{equation}
Substituting this expansion Eq.(\ref{LGseries}) into Eq.(\ref{Paxa}), we have 
\begin{equation}\label{peq}
i\frac{d\xi_{n,l}^{\pm}(z)}{dz} = \sum_{m,k} \langle n,l\vert H_{\pm}\vert m,k\rangle\xi_{mk}^{\pm}
\end{equation}
where 
\begin{equation}
H_{\pm} \approx  i\bigl[Q_{-1}\nabla_{+}+Q_{+1}\nabla_{-}\bigr] \pm \frac{1}{2k_z} [Q_{-2}{\nabla}_{+}^2 - Q_{+2}{\nabla}_{-}^2]
\end{equation}
in which only linear terms in $O(h_{ij})$ are kept.  $\xi_{n,l}^{\pm}(z)$ can be obtained by direct integrating over the summation in Eqn.(\ref{peq}) along the propagating distance $z$.

{\noindent\bf Dipole Interaction - } Let us consider an incident beam of the Hermite-Gauss mode $HG_{10}$, which can be written by a superposition of two LG modes with the opposite OAMs, $l=\pm 1$, 
$$|\hbox{in}\rangle  =  HG_{10} = \frac{1}{\sqrt{2}}(|0,+1\rangle_{LG}+|0,-1\rangle)_{LG} $$
The linear perturbation analysis yields the extra four modes due to the dipole interaction, 
\begin{eqnarray}\label{HGout}
|\hbox{out}\rangle = &|\hbox{in}\rangle + \sqrt{2} Q_1k_wL\cos\gamma\bigl[|0,0\rangle_{LG}+|1,0\rangle_{LG}\bigr] \\ 
&-Q_1 k_wL\bigl[e^{-i\gamma}|0,+2\rangle_{LG} + e^{i\gamma}|0,-2\rangle_{LG} \bigr]\nonumber
\end{eqnarray} 
with
\begin{eqnarray}
&Q_1=\displaystyle{\frac{1}{\sqrt{2}}}\sin\theta_g \sqrt{h_{+}^2\cos^2\theta_g + h_{\times}^2}, \\
&\tan\beta = \displaystyle{\frac{h_{+}}{h_{\times}}}\cos\theta_g, \quad \gamma=\phi_g-\beta \label{ang}
\end{eqnarray} 
where $L$ is the optical length to detector. For each LG$_0^{\pm 1}$ mode,  it carry $\hbar$ of OAM per photon and has a well-known donut-like intensity profile.  Due to the dipole interaction, the induced mode involves one by lowering the OAM of one unit, and consequently gives rise to the vortex-free Gaussian-like beams of $\vert 0,0\rangle_{LG}$ and $\vert 0,1\rangle_{LG}$ modes, producing a bright spot in the center. There are other two high-order modes with $l = \pm 2$,  each acquiring an extra phase factor but with opposite signs, and thus making a global rotation of the intensity pattern by an angle $\alpha$ in Eq.(\ref{ang}).  It is noted that the rotation angle made by the gravitational wave depends on the strain ratio of the cross and plus components - a macroscopic quantity which can be recognized easily if the corresponding signals resulting from GWs are detectable. 

In Fig.(1), we illustrate the central intensity brightening and pattern rotation resulting from the dipole effect with an ideal numerical experiment that displays various intensity patterns in different optical-arm orientations. Experimentally, combing the data obtained from the different photodetectors in this network with the theoretical predictions allows for a more accurate estimation of the GW's physical properties and place more constrains on various gravitational theories.   

\begin{figure}
	\centering
	\includegraphics[width=0.9\textwidth]{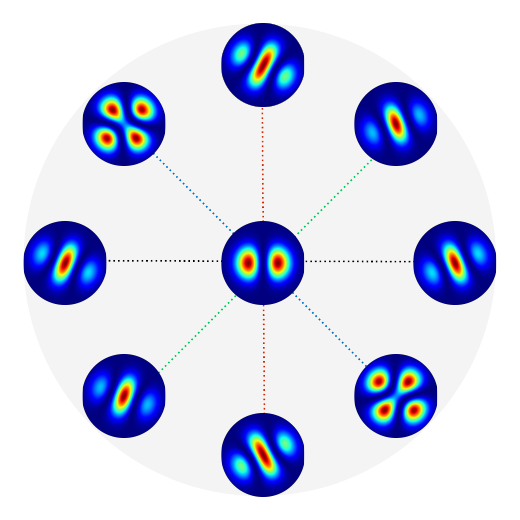} 
	\caption{\label{gw2} Demonstration of the dipole signal in the GW polarization experiment -  We first set up a laboratory coordinate system - the coordinate origin is located at the light source, a network of photodetectors is uniformly placed on the circle centered on the origin, the radius of the circle is L, the xy plane is the plane where the photodetectors are located, and its normal direction points to the zenith, set to the z-axis. In our ideal numerical experiment, we make use of eight detectors, the angle between any two adjacent arms is thus $45^{\circ}$. The split light beams injected from the light source are sent to each photodetectors along the arms. The incident GW is assumed in the direction of $\theta_g=45^{\circ}$, $\phi_g=0^{\circ}$, and the strains is set to $h^{\times}/h^{+}=1$ fixed at $\theta_g=90^{\circ}, \phi_g=0^{\circ}$ in the TT gauge,  Adopting the Hermite-Gauss mode $HG_{10}$, which is a superposition of two LG modes with the opposite OAMs, $l=\pm 1$, $\vert\hbox{in}\rangle  =  HG_{10} = \frac{1}{\sqrt{2}}(\vert 0,+1\rangle_{LG}+\vert 0,-1\rangle)_{LG} $, the additional dipole signal under influence of GWs is composed of the vortex-free mode 
$\sqrt{2}k_wL Q_1\cos\gamma\bigl[\vert 0,0\rangle_{LG} + \vert1,0\rangle_{LG}\bigr]$ and the vortex modes 
$k_wLQ_1 \bigl[e^{-i\gamma}\vert 0,+2\rangle_{LG} + e^{i\gamma}\vert 0,-2\rangle_{LG} \bigr]$
where the complex dipole momentum has been written as $Q_{\pm 1} = Q_1 e^{\pm i\gamma}$. Given the GW's strains and incident direction, $Q_1$ and $\gamma$ vary with the azimuthal angle of optical arms. The diagram presents the false color intensity profile of the input (at the center) and output (the surrounding frames) light beams, indicating both the central brightening and global rotation of intensity patterns. The normalized transverse intensity pattern for an incident HG$_{10}$ mode (upper left),  and output excess mixed modes in different angles of $\alpha= 0^{\circ}$, (upper right) $\alpha = 45^{\circ}$ (lower left) and $\alpha = 90^{\circ}$ (lower right). }
\end{figure}

\begin{figure}[h]%
\centering
\includegraphics[width=0.9\textwidth]{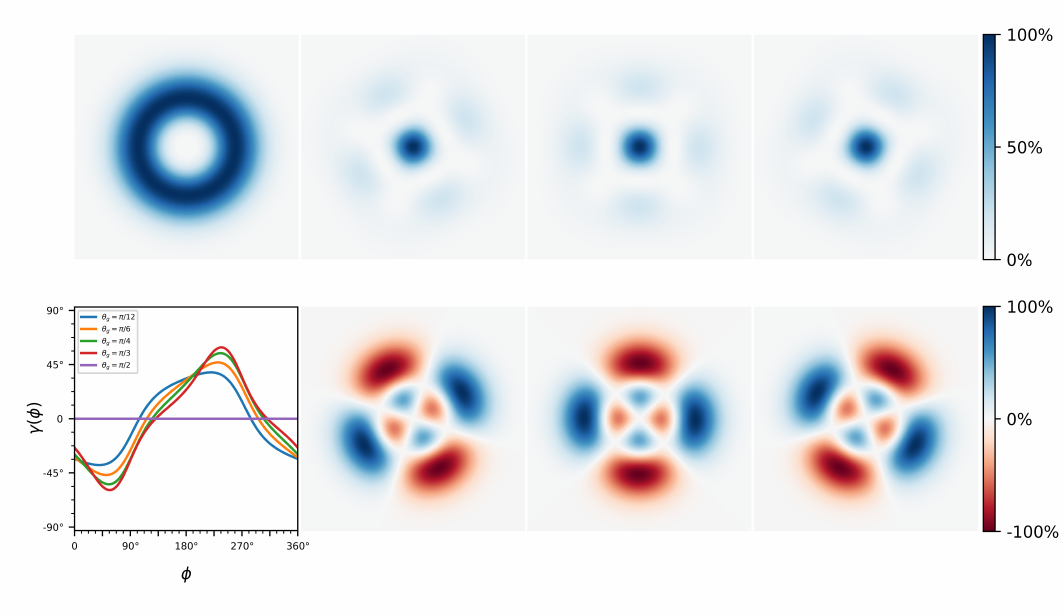}
\caption{The quadrupole signal in the GW polarization experiment -  The laboratory coordinate system is the same as that described in Fig.(1). The incident GW is assumed in the direction of $\theta_g= 30^{\circ}, \phi_g = 0^{\circ}$, and the strains are set to $h^{\times}/h^{+}=1$. The input light beam adopts the LG mode $\vert \hbox{in} \rangle = \vert 0, 2\rangle$, whose transverse intensity profile is plotted in the upper right panel. The quadrupole interaction leads to the extra OAM modes $ \vert \hbox{extra}\rangle = \frac{1}{\sqrt{2}}kL_z Q_2(k_w/k_z)^2\Bigl[e^{-i\gamma}(\vert 0,0\rangle +2\vert 1,0\rangle +  \vert 2,0\rangle) - \sqrt{3} e^{i\gamma}\vert 0,4\rangle\Bigr]$. Given the GW's properties, $Q_2$ and $\gamma$ are functions of the direction angle $\phi$ of optical arms. In the lower left panel, $\gamma(\phi)$ are plotted for different incident polar angles as indicated by the legends. Besides, the right section (2$\times$3 frames) illustrates the false color intensity profiles of the extra modes (upper) and the interference pattern between the input mode and the extra modes (lower), in different azimuthal angles of optical arms, from left to right corresponding to $45^{\circ}$, $135^{\circ}$, $225^{\circ}$ respectively. }\label{fig2}
\end{figure}

{\noindent\bf Quadrupole Interaction - } Since quadrupole interaction is Hermitian conjugate for opposite spin orientations, we consider a superposition of two Laguerre-Gaussian modes with opposite OAMs of quantum number $\pm (l=2)$ and with opposite circular polarizations,  
\begin{equation}\label{Q1}
|\hbox{in}\rangle = F_{l0}(r_{\perp},z)\cdot \Bigl[|{\uparrow}\rangle e^{i2\phi} + |{\downarrow}\rangle e^{-i2\phi}\Bigr]/\sqrt{2}
\end{equation} 
where $|{\uparrow}\rangle=(1,0)^T, |{\downarrow}\rangle=(0,1)^T$ are two 2-vectors for the spin-up and -down states respectively,  and the notation for the LG modes $|n, \pm l\rangle_{LG} = F_{ln}(r_{\perp},z)e^{\pm il\phi}$ have also been used. The input beam expressed by Eq.(\ref{Q1}) is an optical vortex with nonuniform polarization, which can be generated by using liquid-crystal converters \cite{Stalder:96} or spatially varying dielectric gratings\cite{Bomzon:02}.  Applying the linear perturbation analysis, we can find an extra vortex-free optical field produced by the GW's tensor modes,
\begin{eqnarray}\label{Q2}
|\hbox{out}\rangle = & |\hbox{in}\rangle + \displaystyle{\frac{1}{\sqrt{2}}}Q_2 k_zL \Bigl(\displaystyle{\frac{k_w}{k_z}}\Bigr)^2 (F_{00}+2F_{01}+F_{02}) \Bigl[|{\uparrow}\rangle e^{i\gamma} + |{\downarrow}\rangle e^{-i\gamma}\Bigr] \nonumber\\
 & - \sqrt{3}Q_2 k_zL \Bigl(\displaystyle{\frac{k_w}{k_z}}\Bigr)^2F_{04}\Bigl[|{\uparrow}\rangle e^{i(\gamma+4\phi)} + |{\downarrow}\rangle e^{-i(\gamma+4\phi)}\Bigr]
\end{eqnarray}
where 
\begin{eqnarray}
& Q_2 = \sqrt{ h_{+}^{2}(1+\cos^2\theta_g)^2/4 + h_{\times}^2 \cos^2\theta_g} \nonumber \\
& \tan\delta = \displaystyle{ \frac{2\tan\beta}{ (h_+/h_{\times})^2 +\tan^2\beta}},  \quad  \gamma = 2\phi_g + \delta 
\end{eqnarray}
where $\beta$ is given by Eq.(\ref{ang}). As indicated in Eq.(\ref{Q2}), the output beam includes an extra term due to the quadrupole interaction. In the result, the new mode becomes a linear polarization beam with a relative rotation angle $\gamma$, and has a Gaussian-like transverse intensity profile, exhibiting both the central intensity brightening and changes of polarization states. Similar to the dipole effect, the output modes also include high-order modes with $l = \pm 4$, and the corresponding intensity pattern will be rotated by an angle $\gamma$ that depends on both the GW's parameters and the direction angle of lights.  We demonstrate the quadrupole effect in Fig.(2),  where the input light field is assumed to be a single mode $\vert 0, 2\rangle$ with  right-circular polarization.    

Since the SAM of photons keeps unchanged during the quadrupole transition, the changes of OAMs can be attributed to the helicity-2 gravitons. Taking account of  the $\mathcal{T}$ breaking corresponding to the optical gain or loss,  the net effect of quadrupole process can be interpreted by either emitting or absorbing on-shell gravitons. Given the extreme weakness of gravity, the momentum or energy transfer is hardly measured by the available technology, however, the precision quantum optical measurements make it possible to detect the AM transfer from gravitons to photons, and hopefully unveil the quantum nature of gravitation. 

\section{Discussion}

We summary the paper and make some concluding remarks as follows. Based on the four-vector optical Dirac equation developed by Feng \& Wu\cite{2022arXiv220314664F}, we investigate a twisted light beam propagating in gravitational waves. We found that the spin-orbit coupling effect between GWs and structured lights will make photons undergo both the dipole and quadrupole transitions between different orbital-angular-momentum(OAM) eigenstates, the former being spin-independent and the latter spin-dependent. These coupling effects will produce some measurable optical features in the 2-dimensional intensity distribution of twisted light, significantly, the central intensity brightening and macroscopic rotation of the intensity pattern. Moreover, for the spin-dependent quadrupole interaction, it can make the twisted photons with the opposite OAMs and SAMs lose their OAMs and produces a linearly polarized Gaussian light beam, which can be regarded as an alternative spin-orbital Hall effect of lights. It offers a new possible way to realize precision measurements of the gravitational waves, and enables us to explore the polarization modes in 2-dimensional intensity space and extract more information about the physical properties of gravitational waves than the current interferometry experiments.

This paper is focusing on gravitational waves in the Einstein gravity, which only have two polarization modes - cross and plus modes. Obviously, our method can be generalized directly to extended metric gravity theories (e.g. \cite{10.1103/physrev.124.925, tensogravity2008, ROSEN1974455, PhysRevD.8.3293, deRham2014}), in which, there are additional polarization modes including breathing, longitudinal, vector-x, and vector-y. Depending on various incident-angle responses, these modes will exhibit different features in the 2-dimensional intensity distributions, which will allow us to test and discriminate between a variety of gravity theories.    
 
Finally, we note that the newly found dipole and quadrupole interactions, only exist beyond the plane wave approximation and have been omitted in the previous studies, e.g. \cite{Harte_2015, Interferometer2017}. Therefore, including these coupling effects must be important for quantifying precisely the wave behaviors of the higher-order LG modes in optical cavities for the next generation of GW detectors. As demonstrated in this paper, our approach based on the four-vector optical Dirac equation could provide a first-principle field theory method to study the vector-wave mechanics of structured lights in gravitational fields and can be extended to even more generic optical media. 

\authorcontributions{ Conceptualization, F.L.; methodology, F.L. and W.Q.; software, W.Q. and Z.W.; validation, Z.W.; formal analysis, F.L. and W.Q.; investigation, F.L. W.Q. and Z.W.; writing---original draft preparation, F.L.; writing---review and editing, F.L. and Z.W.; visualization, W.Q.; supervision, F.L.; project administration, F.L.; funding acquisition, F.L. and Z.W.  All authors have read and agreed to the published version of the manuscript.}

\funding{This work was supported by the National Key R\&D Program of China through grant 2020YFC2201400 and the Key Program of NFSC through grant 11733010, 11333008, 12173102}

\institutionalreview{Not applicable}

\informedconsent{Not applicable}

\dataavailability{Not applicable} 

\conflictsofinterest{The authors declare no conflict of interest}

\begin{adjustwidth}{-\extralength}{0cm}

\reftitle{References}

\end{adjustwidth}
\end{document}